\documentclass[aps,pre,twocolumn,showpacs,longbibliography,titlepage,superscriptaddress]{revtex4-1}
\usepackage[pdftex]{graphicx}
\usepackage[pdftex,bookmarks,colorlinks,breaklinks]{hyperref}  
\hypersetup{linkcolor=red,citecolor=blue,filecolor=dullmagenta,urlcolor=blue} 
\usepackage{dsfont}

\newcommand{\eqref}[1]{(\ref{#1})}
\newcommand{\bra}[1]{\langle #1 |}
\newcommand{\ket}[1]{| #1 \rangle}
\newcommand{\bk}[2]{\langle #1 | #2 \rangle}
\newcommand{\av}[1]{\left\langle #1 \right\rangle}
\newcommand{\ccdot}{\cdot\,}

\newcommand{\lvert}{|}
\newcommand{\abs}[1]{ \lvert #1 \lvert }	
\newcommand{\modulc}[1]{ \lvert\lvert #1 \lvert\lvert ^{2}}
\newcommand{\modul}[1]{\left\lvert #1 \right\lvert }
\newcommand{\norme}[1]{\lvert \lvert #1 \lvert \lvert }
\newcommand{\normec}[1]{\norme{#1}^{^2}}

\newcommand{\Imm}[1]{\textrm{Im}\left [ #1 \right ]}
\newcommand{\Ree}[1]{\textrm{Re} \left [ #1 \right ]}
\newcommand{\Heff}{H_\textrm{eff}}

\newcommand{\En}{\vec{\cal{E}}_{n} }
\newcommand{\Em}{\vec{\cal{E}}_{m} }
\newcommand{\Enc}[1]{{\cal{E}}_{n,#1 } }
\newcommand{\Ic}{{\cal{I}}}
\newcommand{\Icn}{\Ic_n}

\newcommand{\ro}{\vec{r}_{0} }
\newcommand{\ver}{\vec{r} }

\newcommand{\E}{\vec{E}}
\newcommand{\Gd}{\overline{\overline{G}}}
\newcommand{\1}{\mathds{1}}

\newcommand{\PIvec} {P_{\norme{\E}^{^2}}}
\newcommand{\curl}[1]{\vec{\nabla} \times #1} 

\begin{document}
\title{Lossy chaotic electromagnetic reverberation chambers:\\{U}niversal statistical behavior of the vectorial field}
\author{J.-B.~Gros}\email{jean-baptiste.gros@unice.fr}\thanks{Current Address: ISAE, Universit\'{e} de Toulouse, 10 Av.~E.~Belin BP 54032, 31055 Toulouse, France}
\affiliation{Universit\'e Nice Sophia Antipolis, CNRS, Laboratoire de Physique de la Mati\`ere Condens\'ee, UMR 7336 Parc Valrose, 06100 Nice, France.}
\affiliation{LUNAM Universit\'e, Universit\'e du Maine, CNRS, LAUM, UMR 6613, Av.~O.~Messiaen, 72085 Le Mans, France.}
\author{U.~Kuhl}\email{ulrich.kuhl@unice.fr}
\affiliation{Universit\'e Nice Sophia Antipolis, CNRS, Laboratoire de Physique de la Mati\`ere Condens\'ee, UMR 7336 Parc Valrose, 06100 Nice, France.}
\author{O.~Legrand}\email{olivier.legrand@unice.fr}
\affiliation{Universit\'e Nice Sophia Antipolis, CNRS, Laboratoire de Physique de la Mati\`ere Condens\'ee, UMR 7336 Parc Valrose, 06100 Nice, France.}
\author{F.~Mortessagne}\email{fabrice.mortessagne@unice.fr}
\affiliation{Universit\'e Nice Sophia Antipolis, CNRS, Laboratoire de Physique de la Mati\`ere Condens\'ee, UMR 7336 Parc Valrose, 06100 Nice, France.}
\begin{abstract}
The effective Hamiltonian formalism is extended to vectorial electromagnetic waves in order to describe statistical properties of the field in reverberation chambers. The latter are commonly used in electromagnetic compatibility tests. As a first step, the distribution of wave intensities in chaotic systems with varying opening in the weak coupling limit for scalar quantum waves is derived by means of random matrix theory.
In this limit the only parameters are the modal overlap and the number of open channels.
Using the extended effective Hamiltonian, we describe the intensity statistics of the vectorial electromagnetic eigenmodes of lossy reverberation chambers.
Finally, the typical quantity of interest in such chambers, namely,
the distribution of the electromagnetic response, is discussed.
By determining the distribution of the phase rigidity, describing the coupling to the environment, using random matrix numerical data, we find good agreement between the theoretical prediction and numerical calculations of the response.
\end{abstract}

\maketitle

\section{Introduction}\label{sec:intro}
For more than 40 years, wave chaos has been an attractive field of fundamental research concerning a wide variety of physical systems such as quantum physics \cite{Houches1989}, room acoustics \cite{Mortessagne1993} or ocean acoustics \cite{Tomsovic2010}, guided-wave optics \cite{Pre_Doya}, microwave cavities \cite{Stockmann1990,Stein1992}, \emph{etc}.
The success of wave chaos is mainly due to its ability to describe such a variety of complex systems through a unique formalism which permits us to derive a universal statistical behavior.
Since the early 1990's, quasi bi-dimensional chaotic electromagnetic (EM) cavities have been one of the most used experimental set-ups where these statistical properties have been verified \cite{stoeckmann1999quantum,kuh13}.
In a more applied context, EM cavities are nowadays currently used to realize electromagnetic compatibility (EMC) testing on devices with embedded electronic components \cite{Standard}.
In the EMC community, EM cavities are called EM reverberation chambers (RCs).
Thanks to mechanical \cite{Standard,Leferink2000,Hill_2} or electronic stirring \cite{Standard,Hill_2,Yu2003,Carlberg2009}, enabling ensemble averaging, and to the presence of losses leading to modal overlap, the systems under test are submitted to a supposedly statistically isotropic, uniform, and depolarized electromagnetic field.
The statistical description of the EM field inside RCs commonly used by the EMC community was originally proposed by Hill \cite{Hill1998}, who made a \emph{continuous plane wave spectrum hypothesis}.
This hypothesis assumes that the field is statistically equivalent to a random superposition of traveling plane waves.
This hypothesis is generally well verified if the excitation frequency is much larger than the so-called lowest usable frequency (LUF), the latter being commonly considered to lie between three to six times the fundamental frequency \cite{Bruns2005}.
However, in a frequency regime close to the LUF, the EM field might be neither uniform nor isotropic and this even with stirring \cite{Lunden2007,Arnaut2001}.
In this regime, conventional RCs display a highly non universal behavior depending on their geometry \cite{Bruns2005,Hong2010}, the kind of antennas used \cite{Lemoine2007}, or the object under test \cite{Johnson1998}.

Recent investigations of RCs have concentrated on two goals: (a) improving the RCs' behavior for a frequency regime close to the LUF and (b) finding a suitable statistical model for the EM field in this regime \cite{Arnaut2001,Arnaut2002,Orjubin2006,Lemoine2009} with the incentive of proposing more accurate quantities \cite{Lemoine2009} than those proposed by the IEC standard \cite{Standard}.
In the case of (a), the EMC community has mainly focused on the optimization of the stirring \cite{Otterskog2005,Clegg2005,Weeks2000,Wellander2006,Hong2010} or on the new design of the geometry of RCs \cite{Godfrey1999,Orjubin2009,coro}.
None of these approaches are used to propose quantitative hypotheses concerning the statistics of the EM field near the LUF.
In case of (b), the proposed statistical descriptions converge to Hill's hypothesis in the high frequency limit but are hardly justified on physical grounds.

\begin{figure}
\resizebox{0.6\columnwidth}{!}{\includegraphics{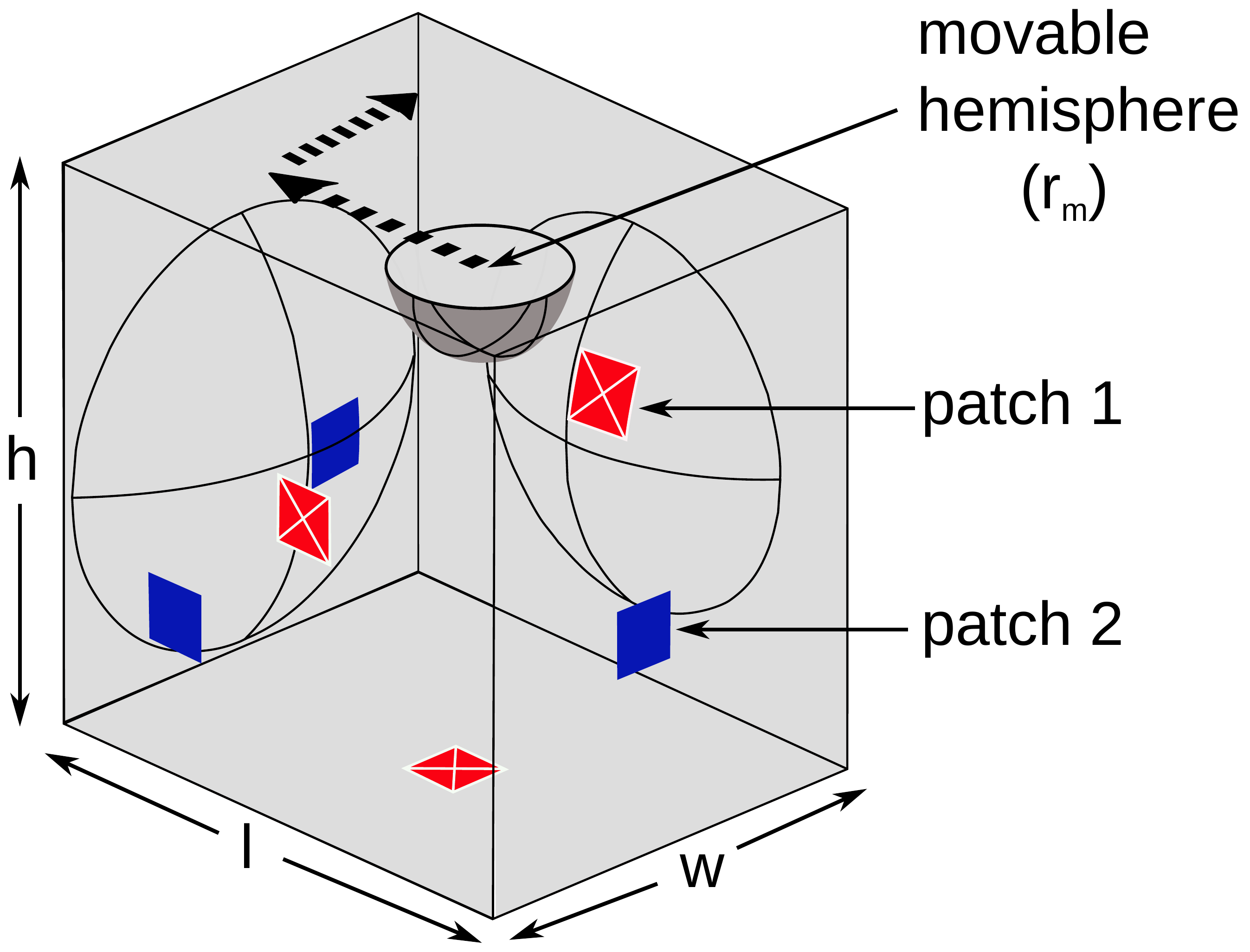}}
\caption{\label{fig:lossy:cav}(color online)
Lossy parallelepipedic reverberation chamber with length $l$=0.985\,m, width $w$=0.785\,m, and height $h$=0.995\,m.
It is made chaotic through the introduction of a movable half-sphere with radius $r_m$=0.15\,m and two spherical caps  with radii of 55\,cm and 45\,cm\cite{GrosWamot2014}.
Losses are mainly localized on sub-surfaces of the walls with a reduced conductivity $\sigma^c_\textrm{abs}$ (colored patches).
The total area of these sub-surfaces is denoted $S_\textrm{abs}$.
There are two configurations, namely, \emph{cavity 1} where only the light red crossed patches (3$\times$patch 1) are attached and \emph{cavity 2} where all six patches are attached (3$\times$patch1 and 3$\times$patch 2).}
\end{figure}

Here we suggest a model based on random matrix theory (RMT) for open chaotic systems to improve the present situation.
Indeed, electromagnetic cavities, and more particularly EM reverberation chambers are genuine examples of open wave systems due, for instance, to Ohmic losses or antenna couplings.
The openness of such cavities turns the real discrete spectrum of closed systems, associated to real eigenfields, into a set of complex resonances.
The latter manifests themselves as complex poles of the scattering matrix which are given by the eigenvalues of the effective non-Hermitian Hamiltonian $\Heff$ of the open system \cite{verb85,soko89,fyod97,rott09}.
The anti-Hermitian part of $\Heff$ arises from coupling between the internal (bound) and continuum states, leading to finite resonance widths.
The other key feature is that the eigenfunctions of $\Heff$ are nonorthogonal \cite{soko89,rott09,gro14}.
For systems invariant under time reversal, like the open EM cavities studied below (see Fig.~\ref{fig:lossy:cav}), the nonorthogonality manifests itself in complex wave functions, yielding the phase rigidity \cite{brou03,kim05,bulg06b} and mode complexness \cite{savi06b,poli09b,gro14}.
In practice, the EM response of such open systems consists of a sum of overlapping resonances, associated to complex eigenfields (the resonant or decaying states) which can be viewed as a sum of standing and traveling waves \cite{Pnini}.

In this paper we will show how the $\Heff$ approach \cite{kuh13} allows to extend the random matrix theory to fully three-dimensional (3D) open chaotic EM reverberation chambers.
Hence, universal statistical spectral and spatial behaviors can be predicted for these systems.
This has been recently illustrated in Ref.~\cite{gro14}, where the experimental width shift distribution was investigated in a chaotic reverberation chamber as a test of non-orthogonality.
We will show here that these crucial properties for electromagnetic compatibility tests, especially the field uniformity, can indeed be achieved in chaotic EM cavities \cite{GrosWamot2014} even at low enough frequency where the modal overlap is only weak or moderate.
In a conventional RC with a regular rectangular shape, these low-lying modes display highly non-isotropic patterns resulting in a low-frequency response which cannot be expected to be statistically isotropic and uniform, even in the presence of stirring.
To the contrary, these statistical requirements are naturally fulfilled by the vast majority of modes in a chaotic cavity without the help of any stirring process \cite{GrosWamot2014}.
This generic statistical behavior of modes in a chaotic cavity will be referred to as \emph{ergodicity} in the following.
In particular, we will show how the ergodicity of modes in a chaotic RC permits to establish a statistically uniform response.
This behavior is in complete contradiction with what can be obtained in a conventional RC where the specific details of the chamber lead to unpredictable statistical features \cite{GROS2014AEM,Gros2015MAS}.

In the following section, we introduce important notations and remind the reader of the main universal properties of open chaotic systems, in particular concerning the spatial distribution of complex eigenstates.
Most are obtained through the effective Hamiltonian formalism and have been successfully verified experimentally and numerically in quasi-2D lossy chaotic cavities (see, for instance, Ref.~\cite{kuh13} and references therein).
In the third section, we use the Dyadic Green's function representation of the response of a fully 3D vectorial EM cavity to extend the previous results by properly defining the bi-orthogonality of complex modes for vectorial fields and their phase rigidity.
Then, we establish the universal statistics of the complex modes and of the full response in chaotic RCs.
We successfully compare all the predictions deduced from our extended RMT approach to numerical simulations of realistic RCs.
We conclude by showing how a minimal number of ingredients, namely, the number and the coupling strength of fictitious equivalent channels, is sufficient to give a full statistical account of the intensity distributions that can be measured in a chaotic RC, irrespective of the specific details of the cavity shape and of the antenna set-up.

\section{Chaotic open system: universal statistics with $\Heff$ formalism}\label{sec:heffForm}
The response of an open scalar wave system can be formally recovered by means of the Green's functions formalism used in the quantum context for open systems, through \cite{soko92}
\begin{equation}\label{Green_mq}
  G(\epsilon)=(\epsilon-\Heff)^{-1},
\end{equation}
where $\Heff$ is the (non-Hermitian) effective Hamiltonian and is commonly written as \cite{soko92,rott09,oko03}
\begin{equation}\label{eq:Heff:def}
\Heff = H-\frac{i}{2}W,
\end{equation}
where $H$ is the Hermitian part corresponding to the closed system (neglecting the Hermitian effect due to opening) and $-\frac{i}{2}W$ is the anti-Hermitian part related to coupling to the environment.
In systems with preserved time reversal symmetry (which is assumed for EM cavities \cite{stoeckmann1999quantum,kuh13}), $H$ is a real symmetric operator. We assume $W$ to be real as well since any imaginary part induced by the coupling can be included in $H$\cite{soko89,koeb10,kuh13}.
The complex eigenvalues of $\Heff$, $\epsilon_n=\tilde{\epsilon}_n-i\Gamma^\epsilon_n/2 $, yield the center $\tilde{\epsilon}_n$ of the $n$th resonance and its width $\Gamma^\epsilon_n$.
As a consequence of the non-Hermiticity of $\Heff$, there are two distinct sets of complex left $\left\lbrace \bra{L_n}\right\rbrace$ and right $\left\lbrace \ket{R_n}\right\rbrace$ eigenvectors, associated to the same set of eigenvalues $\left\lbrace\epsilon_n\right\rbrace$:
\begin{equation}
\bra{L_n} \Heff= \epsilon_n \bra{L_n} \quad \textrm{and} \quad \Heff \ket{R_n}=\epsilon_n \ket{R_n}.
\end{equation}
The eigenvectors $\bra{L_n}$ and $\ket{R_n}$, which describe the resonance states, form a complete and bi-orthogonal set:
\begin{eqnarray}
&\sum_n \ket{R_n}\bra{L_n}=1, \\
&\bk{L_n}{R_m}=\delta_{nm}. \label{n_and_b}
\end{eqnarray}
The time reversal symmetry assumption adds another constraint on right and left eigenvectors \cite{rott09,soko89,poli09b}, namely that they are the transpose vector of each other:
\begin{equation}\label{r_et_l}
\bra{L_n}=\ket{R_n}^T.
\end{equation}
Note that the bi-orthogonality condition (\ref{n_and_b}) implies that the real and imaginary parts of the wave functions are orthogonal to one another.
The Green's function (\ref{Green_mq}) can be expanded on the left and right eigenvectors \cite{oko03,rott09}:
\begin{equation}\label{Green_mq_exp}
G(\epsilon)=\sum_n \frac{\ket{R_n}\bra{L_n}}{\epsilon-\epsilon_n}.
\end{equation}
For the present study, important quantities are the mean level spacing $\Delta$ and the local average resonance width ${\Gamma^\epsilon}$, whose ratio permits to evaluate the amount of overlap between adjacent resonances through the mean \emph{modal overlap}:
\begin{equation}\label{def_d}
d=\frac{{\Gamma^\epsilon}}{\Delta}.
\end{equation}

\subsection{$\Heff$ model for chaotic cavity and RMT results}
RMT is a powerful tool to derive universal statistical properties of closed \cite{BGS,stoeckmann1999quantum,Guhr1998} and open wave chaotic systems \cite{kuh13,savi06b,Poli2010,Poli2012,Poli2009_avoid,Sommers1999,Kuhl2005}.
In this framework, the effective Hamiltonian of an open cavity with $N$ resonances is modelled by an $N \times N$ random matrix and loss mechanisms are associated with $M$ open channels connecting the $N$ levels of the closed (lossless) cavity to its environment.
As we have in mind to study the statistics of the vectorial field of actual chaotic reverberation chambers, we concentrate on systems which preserve time reversal symmetry.
Then the Hamiltonian of the closed system, $H$ in Eq.~\eqref{eq:Heff:def}, is modelled by a random matrix belonging to the Gaussian Orthogonal Ensemble (GOE) \cite{stoeckmann1999quantum,BGS}.
The anti-Hermitian part $-\frac{i}{2}W$ of $\Heff$ is related to a $N\times M$ coupling matrix $V$ through $W=VV^\dagger$.
Each element $V^j_n$ of $V$ connects the $n$-th eigenstate of $H$ ($n \in \left[ 1,N\right] $) to the $j$-th open channel ($j \in \left[ 1,M\right] $).
In order to preserve orthogonal invariance of $\Heff$ under orthogonal transformations \cite{soko89}, the $V^j_n$ are commonly set as real independent Gaussian random variables with zero mean and standard deviation $\sigma_j$ depending only on channel $j$ (as a consequence of the ergodicity of eigenstates) and related to its coupling strength $\kappa_j$ through \cite{poli09b,savi06b,soko89}
\begin{equation}
\sigma_j^2=(2\kappa_j\Delta)/\pi. \label{eq:sigma_v}
\end{equation}
In the following, we will consider equivalent channels associated to Ohmic losses and will therefore assume that all channels are identically coupled \cite{savi06b,poli09b}:
\begin{equation}\label{Stat_V}
\av{V^j_n V^{k}_{p}}=\sigma^2 \delta^{jk} \delta_{np}= (2\kappa\Delta/\pi) \delta^{jk} \delta_{np}.
\end{equation}

We will now establish a few results, obtained in the framework of RMT, concerning the statistical features of scalar eigenmodes for chaotic open wave systems, as verified through experiments in 2D chaotic microwaves cavities, and see, in Sec.~\ref{sec:response}, how they can be extended to the case of vectorial fields.

\subsection{Spatial distribution of the intensity of eigenstates}
For scalar fields, eigenmodes of open chaotic systems are complex wave functions $\Psi_n(\vec{r})=\bk{\vec{r}\,}{R_n}$, whose spatial statistics can be described by the statistics of the $N$ components $\Psi_{n,i}$ ($i=1,\cdots,N$) of the right eigenvectors of the random matrix $\Heff$.
The bi-orthogonality condition (\ref{n_and_b}) and GOE assumption imply:
\begin{eqnarray}
&&\sum_{i=1}^N \Ree{\Psi_{n,i}}\Imm{\Psi_{n,i}}=0, \label{IR_orthog}\\
&&\sum_{i=1}^N \left(\Ree{\Psi_{n,i}}^2-\Imm{\Psi_{n,i}}^2\right)=1, \label{RmoinsI_norme} \\
&&\sum_{i=1}^N \left(\Ree{\Psi_{n,i}}^2+\Imm{\Psi_{n,i}}^2\right)\geq 1, \label{RplusI}
\end{eqnarray}
where $\Ree{\Psi_{n,i}}$ and $\Imm{\Psi_{n,i}}$ are two independent Gaussian distributed random variables.
It is important to note that $\Ree{\Psi_{n,i}}$ and $\Imm{\Psi_{n,i}}$ do not necessarily have the same variances.
The complexness parameter, $q^2_n$, originally introduced by Lobkis and Weaver in Ref.~\cite{Lobkis2000}, is defined by the ratio of the variances of the imaginary and real parts of $\Psi_n$ for systems which preserve time reversal symmetry \cite{Barthelemy2005_2,savi06b,poli09b} :
\begin{equation}\label{q2_scal}
  q_n^2=\frac{\sum_{i=1}^N \Imm{\Psi_{n,i}}^2}{\sum_{i=1}^N \Ree{\Psi_{n,i}}^2}.
\end{equation}
This parameter allows to quantify the sensitivity of the $n$th eigenmode to the openness of the system.
It is related to the \emph{phase rigidity} $\abs{\rho_n}^2$ \cite{brou03,kim05,bulg06b}, where ${\rho_n= \sum_{i=1}^N {\Psi^2_{n,i}}/\sum_{i=1}^N \abs{\Psi_{n,i}}^2}$, through:
\begin{equation}\label{q2_et_rho}
  q_n^2=\frac{1-\modul{\rho_n}}{1+\modul{\rho_n}}\,.
\end{equation}
Using the above mentioned statistical assumptions, the spatial distribution of the normalized intensity $\tilde{I}_n=\abs{\Psi_n}^2/\av{\abs{\Psi_n}^2}$ of a complex eigenstate (where $\av{..}$ corresponds to a spatial average) reads
\begin{equation}\label{P_de_In}
  P(\tilde{I}_n;\rho_n)=\frac{1}{\sqrt{1-\abs{\rho_n}^2}} \exp\left[ -\frac{\tilde{I}_n}{1-\abs{\rho_n}^2} \right] \textrm{I}_0\left[\frac{\abs{\rho_n}\tilde{I}_n}{1-\abs{\rho_n}^2} \right]\,.
\end{equation}
This distribution interpolates between the Porter-Thomas distribution ($\modul{\rho}=1$) for closed systems and the Rayleigh distribution ($\modul{\rho}=0$) for fully open systems \cite{Ishio2001}.
It was formerly proposed by Pnini and Shapiro who considered the complex wave field as a superposition of standing and traveling plane waves \cite{Pnini}.
This distribution had also been proposed previously by \.Zyczkowski and Lenz in the different context of the crossover induced by time reversal breaking in chaotic systems \cite{Zyczkowski1991}.
Strangely enough, to our knowledge, it was never shown to be valid for the complex eigenvectors of the non-Hermitian random Hamiltonian $\Heff$.
Since $q_n$ (hence $\rho_n$) was demonstrated to be a distributed quantity \cite{Poli2009_t,poli09b,Poli2010}, the distribution of the normalized intensity of eigenstates over the statistical ensemble is given by
\begin{equation}
 P(\tilde{I})=\int p_\rho(\rho) P(\tilde{I}; \rho) d\rho. \label{P_de_In_global}
\end{equation}
In the weak coupling regime defined by:
\begin{equation}\label{weak_coupling}
\sqrt{\textrm{Var}(\Gamma_n)}/\Delta \ll 1,
\end{equation}
and for small or moderate modal overlap $d$, the distribution of the complexness parameter $q_n$ of chaotic open wave system has been derived in Refs.~\citep{Poli2009_t,poli09b} by using perturbation theory, and reads
\begin{equation}\label{p_de_x}
p_{x;M}(x_n)=\frac{\pi ^2 \left(1+\frac{(3+M) \pi ^2}{4 M^2 x_n}\right)}{24 M x_n^2 \left(1+\frac{\pi ^2}{4 M^2 x_n}\right)^{2+\frac{M}{2}}},
\end{equation}
where $x_n=q_n^2/d^2$ and $M$ is the number of open channels.
The distribution $p_\rho(\rho)$ follows from $p_\rho(\rho) \, d\rho = p_x(x) \, dx$ and is given by
\begin{equation}
 p_{\rho ;\lbrace M,d\rbrace}(\abs{\rho_n})=\frac{2}{d^2 (1+\abs{\rho_n})^2} \,p_{x;M}\left(\frac{(1-\abs{\rho_n})}{d^{2} (1+\abs{\rho_n})}\right). \label{eq:p:de:rho}
\end{equation}

The number of open channels $M$ is the only parameter of the distribution (\ref{p_de_x}) and is also related to the fluctuations of the widths.
In the weak coupling regime, denoting $\ket{n}$ the $n$th eigenstate of $H$, and by considering the anti-Hermitian part of $\Heff$ as a perturbation of $H$, the perturbation theory gives
\begin{equation}\label{gamma_n_perturbatif}
\Gamma_n^\epsilon =\bra{n} VV^\dagger \ket{n}=\sum_{j=1}^M {V^j_n}^2.
\end{equation}
Thus $\Gamma_n^\epsilon$ corresponds to a sum of squares of independent identically distributed Gaussian random variables with variances $\sigma^2$ [Eq.~\eqref{eq:sigma_v}].
This yields a mean width
\begin{equation}\label{mean_gamma}
  {\Gamma}^\epsilon=\sigma^2 M,
\end{equation}
and a variance
\begin{equation}\label{Mvar_gamma}
\textrm{Var}(\Gamma^\epsilon_n)=2\sigma^4 M,
\end{equation}
and the rescaled widths $\gamma_n=\Gamma_n^\epsilon/\sigma^2$ are chi-square distributed with $M$ degrees of freedom \cite{Poli2009_t,poli09b,Poli2010,gro14}:
\begin{equation}\label{P_de_gamma_n}
  P_{\gamma;M}(\gamma_n)=\frac{1}{2^{M/2}\Gamma(M/2)} \gamma_n^{M/2-1} e^{-\gamma_n/2}.
\end{equation}
Therefore the first two moments of $\Gamma_n^\epsilon$ are directly related to $M$ through
\begin{equation}\label{eq:M:gamma}
  \frac{M}{2}=\frac{\left(\Gamma^\epsilon\right)^2}{\textrm{Var}(\Gamma_n^\epsilon)},
\end{equation}
and the normalized width $\tilde{\Gamma}_n=\Gamma_n^\epsilon/{\Gamma^\epsilon} $, follows the distribution:
\begin{equation}\label{P_de_Gamma}
  P_{\Gamma;M}(\tilde{\Gamma})=\frac{1}{\Gamma(a)b^a} \tilde{\Gamma}^{a-1} e^{-\tilde{\Gamma}/b},
\end{equation}
where $M$ is the only parameter since $a=b^{-1}=M/2$.
It is important to note that the number $M$ and the modal overlap $d$ are essential parameters for open chaotic systems.
Indeed, by combining the expressions (\ref{mean_gamma}) and (\ref{Stat_V}), the coupling strength $\kappa$ can be expressed in terms of these parameters by:
\begin{equation}\label{kappa_d}
  \kappa=\frac{\pi {\Gamma}^\epsilon}{2\Delta M}=\frac{\pi d}{2 M}.
\end{equation}
One can also note that Eq.~\eqref{eq:M:gamma} permits to restate the weak coupling condition (\ref{weak_coupling}) in terms of $M$ and $d$ as
\begin{equation}
  d\sqrt{\frac{2}{M}} \ll 1\,.
\end{equation}

\begin{figure}
  \mbox{
    \raisebox{3.75cm}[0cm][0cm]{(a)}\hspace*{-1.5em}
      {\resizebox{0.49\columnwidth}{!}{\includegraphics{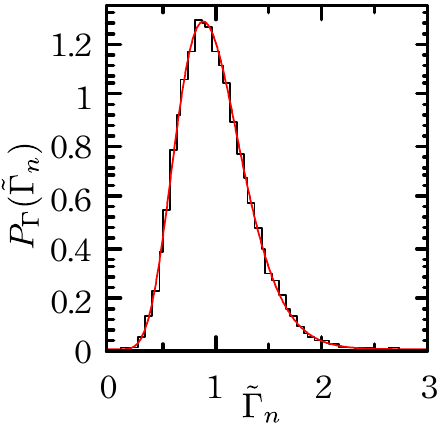}}}
    \raisebox{3.75cm}[0cm][0cm]{(b)}\hspace*{-1.5em}
      {\resizebox{0.49\columnwidth}{!}{\includegraphics{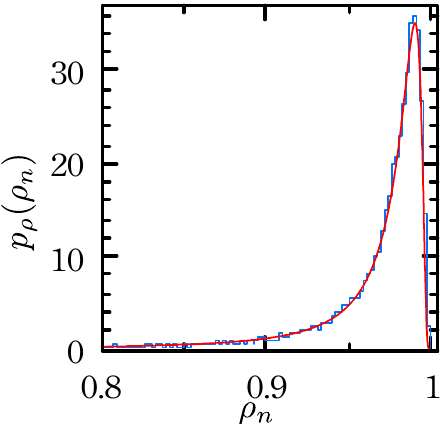}}}}
    \raisebox{5.5cm}[0cm][0cm]{(c)}\hspace*{-2em}
      {\resizebox{!}{!}{\includegraphics{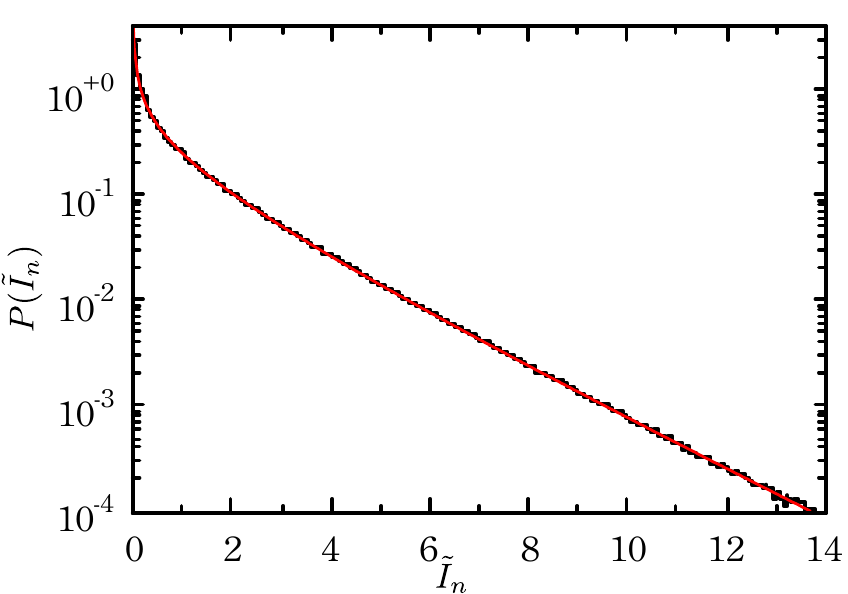}}
  }
\caption{\label{fig:rmt:weak}(color online)
The histogram corresponds to (a) the width distribution, (b) the modal phase rigidity distribution, and (c) the intensity distribution obtained by RMT calculations using} an ensemble of eigenvectors of 150 random $\Heff$ with $N=700$, $M=20$ and $d=0.50$.
The continuous red curves correspond to the weak coupling predictions given by Eq.~\eqref{P_de_Gamma} for (a), Eq.~\eqref{eq:p:de:rho} for (b) and Eq.~\eqref{P_de_In_global} for (c) with $p_\rho= p_{\rho ;\lbrace M,d\rbrace}$.
\end{figure}

\begin{figure}
    \raisebox{3.75cm}[0cm][0cm]{(a)}\hspace*{-1.5em}
      {\resizebox{0.49\columnwidth}{!}{\includegraphics{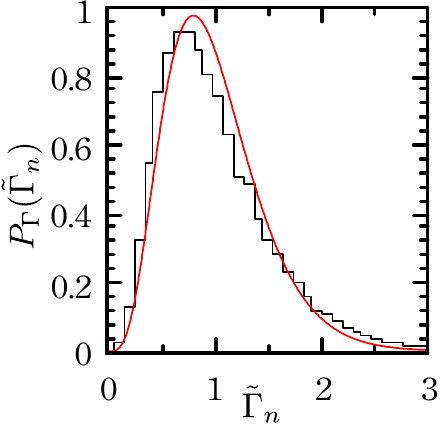}}}
    \raisebox{3.75cm}[0cm][0cm]{(b)}\hspace*{-1.5em}
      {\resizebox{0.49\columnwidth}{!}{\includegraphics{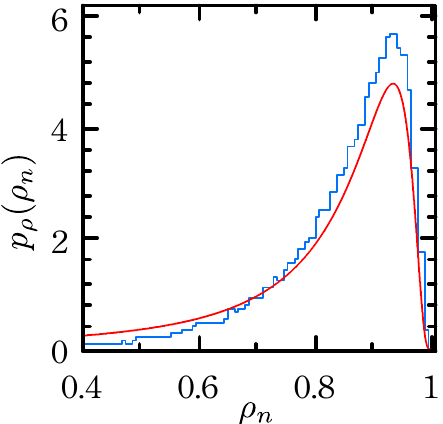}}}
    \raisebox{5.5cm}[0cm][0cm]{(c)}\hspace*{-2em}
      {\resizebox{!}{!}{\includegraphics{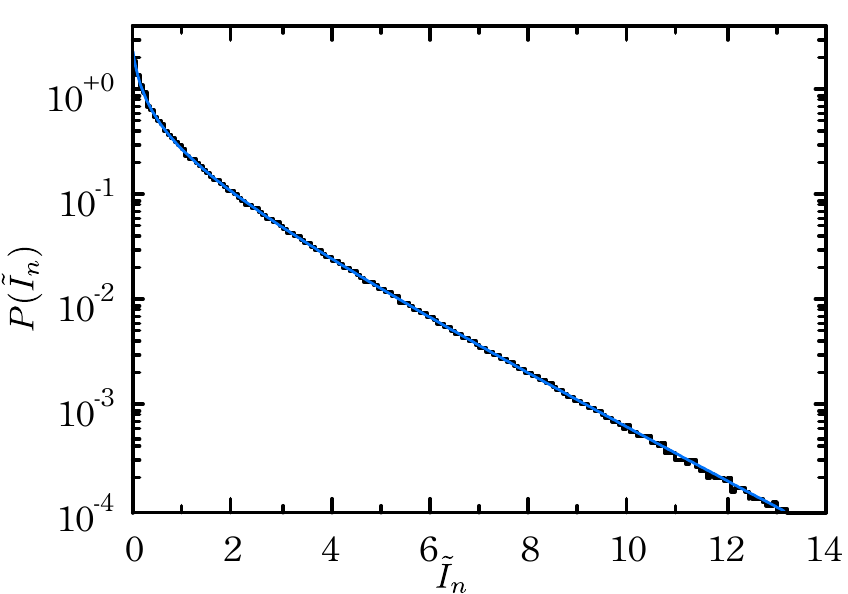}}}
\caption{\label{fig:rmt:non:weak}(color online)
The histogram corresponds to (a) the width distribution, (b) the modal phase rigidity distribution and (c) the intensity distribution obtained by RMT calculations using an ensemble of eigenvectors of 150 random $\Heff$ with $N=700$, $M=10$ and $d=1$.
The continuous red curves correspond to the weak coupling predictions given by Eq.~\eqref{P_de_Gamma} for (a) and Eq.~\eqref{eq:p:de:rho} for (b).
In panel (c), the blue continuous curve is obtained through Eq.~\eqref{P_de_In_global} where $p_\rho$ is given by the discrete empirical distribution [blue histogram shown in (b)].}
\end{figure}

All the above mentioned results are checked and illustrated in Figs.~\ref{fig:rmt:weak} and \ref{fig:rmt:non:weak}.
For each figure, we have diagonalized 150 random $\Heff$ with $N=700$.
We have kept only eigenvalues and their associated eigenvectors in the center of the semi-circular law ($\epsilon\simeq 0$), where one can assume that $\Delta$ is almost constant.
The results shown in Figs.~\ref{fig:rmt:weak} and in \ref{fig:rmt:non:weak} are respectively obtained for two different values of the modal overlap and number of coupling channels, namely $d=0.5$ and $M=20$ and $d=1$ and $M=10$.
The parameter $d\sqrt{2/M}$ is respectively $0.16$ and $0.45$ for the results associated with Figs.~\ref{fig:rmt:weak} and \ref{fig:rmt:non:weak}.
Thus, we can assume \emph{a priori} that the results shown in Fig.~\ref{fig:rmt:weak} must correspond to the weak coupling regime whereas the results shown in Fig.~\ref{fig:rmt:non:weak} will deviate from this regime.
This point is indeed supported by graphs on top where the empirical distributions of normalized width $\tilde{\Gamma}_n$ [black histograms in Figs.~\ref{fig:rmt:weak}(a) and \ref{fig:rmt:non:weak}(a)] and the empirical distributions of modal phase rigidity $\rho_n$ [blue histograms in Figs.~\ref{fig:rmt:weak}(b) and \ref{fig:rmt:non:weak}(b)] are respectively compared with the perturbative predictions (red continuous curves) $P_{\Gamma;M}$ [Eq.~\eqref{P_de_Gamma}] and $ p_{\rho ;\lbrace M,d\rbrace}$ [Eq.~\eqref{eq:p:de:rho}], with $M$ and $d$ taking the values used in RMT simulations (see captions).
Only the results of Fig.~\ref{fig:rmt:weak} tally with the weak coupling assumption.
For Figs.~\ref{fig:rmt:non:weak}(a) and \ref{fig:rmt:non:weak}(b), the visible deviation between the histograms and the continuous curves is due to a larger value of the parameter $d \sqrt{2/M}$ for which a weak coupling assumption is clearly no longer valid.
For completeness, we also compare the exact number of coupling channels with the nearest integer given by expression \eqref{eq:M:gamma} and with the value given by fitting the empirical distribution of $\rho_n$ with $p_{\rho;\lbrace M,d \rbrace}$ where $M$ is left as a free parameter and $d$ fixed by RMT simulation.
For the results corresponding to Fig.~\ref{fig:rmt:non:weak}, there are clear discrepancies between the values returned by the fit $M=13$ and $2\Gamma^2/\textrm{Var}(\Gamma_n)=7$ whereas for the results corresponding to the weak coupling regime (Fig.~\ref{fig:rmt:weak}) the estimation of $M$ through expression \eqref{eq:M:gamma} and by fitting $p_\rho$ are both consistent with $M=20$.
The reader's attention should be drawn to the fact that the different ways of estimating the number of channels [either through Eq.~\eqref{eq:M:gamma} or by fitting $p_\rho$] are indeed mutually consistent as long as the perturbative regime holds and gradually disagree when leaving it.
In Figs.~\ref{fig:rmt:weak}(c) and \ref{fig:rmt:non:weak}(c) we compare the empirical distributions of all normalized intensities of the above mentioned eigenvectors (black histograms) with the formula \eqref{P_de_In_global} [red continuous curve in Fig.~\ref{fig:rmt:weak}(c) and blue continuous curve in Fig.~\ref{fig:rmt:non:weak}(c)].
Irrespective of the coupling regime, a remarkable agreement is observed over more than four orders of magnitude. To the best of our knowledge, this is the first time that the eigenvectors of random matrices associated to $\Heff$ are shown to follow the distribution \eqref{P_de_In}.
However, there is a slight 	difference between the theoretical curve plotted in both figures.
Indeed, the red curve of Fig.~\ref{fig:rmt:weak}(c) has been obtained by substituting in formula \eqref{P_de_In_global} $p_\rho$ for the perturbative $p_{\rho ;\lbrace M,d\rbrace}$ [red curve shown in Fig.~\ref{fig:rmt:weak}(b)], whereas because of the departing from the weak coupling regime, the blue curve of Fig.~\ref{fig:rmt:non:weak}(c) have been built from the empirical $p_\rho$ [blue histogram shown in Fig.~\ref{fig:rmt:weak}(b)].
If in this regime, instead of using the empirical $p_\rho$, we use the $p_{\rho ;\lbrace M,d\rbrace}$ shown in Fig.~\ref{fig:rmt:non:weak}(b), the resulting distribution differs from the blue curve by more than $12\%$ near the most probable values and by more than $20\%$ on the tail.
In both cases, the Porter-Thomas distribution has the wrong tail behavior and deviates more than 10\% from the red and blue curves in Figs.~\ref{fig:rmt:weak}(c) and in \ref{fig:rmt:non:weak}(c), respectively.

\section{Vectorial response in lossy EM 3D cavities}\label{sec:response}
We now consider the EM response in a RC as a continuous function of the excitation frequency $f$.
The electric field, $\E(\ver,f)$ created by an arbitrary current source $\vec{J}_e(\ro,f)$ localized in a volume $V_0$ and oscillating at frequency $f$, is the solution of the vectorial Helmholtz equation \cite{Morse1953,Collin1991}:
\begin{equation}
 \curl{\curl{\E}} - k^2 \E = -2\pi i f \mu_o \vec{J}_e,\label{eq:helm:vec}
\end{equation}
where $k=2 \pi f /c$.
The electric field can be written in term of the convolution of the source term with the so called Dyadic Green's function (DGF), denoted $\Gd(\ver,\ver_0,f)$, \cite{Collin1991,Morse1953}:
\begin{equation}
  \E(\ver,f)=-i\omega\mu_o \int_{V_0} \Gd(\ver,\ver_0,f) \cdot \vec{J}_e(\ver_0,f) d\ver_0,
\end{equation}
Indeed, the DGF is the solution of
\begin{equation}
  \curl{\curl{\Gd(\ver,\ver_0, f)}}- k^2 \Gd(\ver,\ver_0, f) = \1 \delta(\ver-\ver_0) . \label{eq:helm:TGD}
\end{equation}
Since we are only interested in modeling the statistical properties of the EM response, the DGF appears to be an appropriate tool.
The DGF is often written as a matrix,
\begin{equation}\label{FGD_mat}
\overline{\overline {G}}(\vec{r},\vec{r}_0,f)=
\left(
\begin{array}{ccc}
G_{xx} & G_{xy} & G_{xz} \\
G_{yx} & G_{yy} & G_{yz} \\
G_{zx} & G_{zy} & G_{zz} \\
\end{array}\right),
\end{equation}
where each column (associated to the second subscript $j=x$, $y$, or $z$) stands for the three components of the electric field vector at the spatial coordinate $\vec{r}$:
\begin{equation}
\E_j(\vec{r},f)= \Gd(\vec{r}, \vec{r}_0, f)\cdot \vec{e}_j =\left(\begin{array}{c}
G_{xj} \\
G_{yj} \\
G_{zj}
\end{array}\right).\label{eq:E:avec:tgd}
\end{equation}
Thus the vector $\lbrace G_{xj} , G_{yj} , G_{zj} \rbrace$ is the EM response created in the cavity by a point-like elementary current source located at $\vec{r}_0$, polarized along the Cartesian unit vector $\vec{e}_j$, and oscillating at frequency $f$.
Following the formalism given in Refs.~\cite{Collin1991,GrosWamot2014}, the DGF can be also expanded over the resonances to read:
\label{stop}
\begin{equation}\label{Green}
  \overline{\overline {G}}(\vec{r},\vec{r}_0,f)=\sum_{n=1}^{\infty} \frac{ \vec{E}_n(\vec{r})\otimes\vec{E}_n(\vec{r}_0)} { (k_n^2-k^2)},
\end{equation}
where $k=2\pi f/c $.
In the latter expression, losses lead to complex (non-real) eigenmodes (resonance states) $\vec{E}_n$ \cite{Barthelemy2005_2,Poli2010}, associated with complex eigenvalues $k_n$
\begin{equation}
  k_n=\frac{2\pi f_n}{c}\left(1-\frac{i}{2 Q_n}\right),
\end{equation}
where $c$ is the velocity of light, $f_n$ is the central frequency of the $n$th resonance and $Q_n$ is its quality factor related to the $n$th resonance bandwidth $\Gamma^f_n=f_n/Q_n$.
In expression (\ref{Green}), the irrotational contribution has been omitted since it can be neglected far from the source region ($\vec{r}\neq\vec{r}_0$).
The amount of overlap between adjacent resonances can be evaluated through the mean modal overlap, here defined by the ratio of the mean bandwidth ${\Gamma^f}$ and the mean resonant frequency spacing $\Delta^f$:
\begin{equation}
  d= {\Gamma^f}/\Delta^f \,.
\end{equation}
In the following, as we are only concerned with EM waves, we will omit the $f$ superscript.

\subsection{Bi-orthogonality and phase rigidity for vectorial fields}
Equation~(\ref{Green_mq_exp}) can be re-written in the spatial representation as Eq.~(\ref{Green}), which is an extension of the case of the scalar field, where
\begin{equation}\label{eq:GScalar}
G(\vec{r},\vec{r}^\prime,k)=\sum_n\frac{\Psi_n(\vec{r})\Psi_n(\vec{r}^\prime)}{k^2-k_n^2},
\end{equation}
which is a solution, inside of the cavity, of
\begin{equation}\label{eq:HelmScalar}
\left(\Delta_{\vec{r}} + k^2\right)G(\vec{r},\vec{r}^\prime,k)=\delta(\vec{r}-\vec{r}^\prime)
\end{equation}
together with the boundary conditions imposed by the coupling channels \cite{bar05a,Barthelemy2005_2}.
In contrast to the Hermitian problem, where the residues are given by $\Psi_n(\vec{r})\Psi^*_n(\vec{r}^\prime)$, here we have the residues $\Psi_n(\vec{r})\Psi_n(\vec{r}^\prime)$ due to the relation $\bra{L_n}=\ket{R_n}^T$. Note that different sign conventions are used in the quantum case [Eq.~\eqref{eq:GScalar}] and in the electromagnetic case [Eq.~\eqref{Green}].
We wish to draw the reader's attention to the central fact that expressions (\ref{Green}) and (\ref{Green_mq_exp}) can be made formally equivalent provided that the $\vec{E}_n$'s are properly normalized.
This normalization is achieved through the following transformation of the vector field:
\begin{equation}\label{norma_vec}
\En = \vec{E}_n/\sqrt{\int_V\vec{E}_n\cdot\vec{E}_n dV},
\end{equation}
leading to an explicit form of the bi-orthogonal condition for vectorial eigenfields [formally equivalent to (\ref{n_and_b})]:
\begin{equation}\label{norm_int}
\int_V\En \cdot \Em dV= \delta_{nm},
\end{equation}
where $\int_V ...dV$ is the integral over the volume $V$ of the cavity. Note that, in the scalar product, the standard conjugation is omitted due to the constraint (\ref{r_et_l}).
Condition (\ref{norm_int}) was shown to provide a smooth transition from the complex wave functions of an open wave system to the real wave functions of the corresponding closed system as
\begin{equation}
  \int_V \modulc{\En} dV \geq 1,
\end{equation}
and tends to unity for vanishing anti-Hermitian part of the effective Hamiltonian $\Heff$.
To evaluate this transition, the definition of the phase rigidity for vector fields is extended as follows:
\begin{equation}
  \rho_n= \frac{\int_V\vec{E}_n\cdot\vec{E}_n dV}{\int_V\modulc{\vec{E}_n} dV},
\end{equation}
whose modulus reads
\begin{equation}\label{modul_rho_n_vec}
  \modul{\rho_n}=\frac{1}{\int_V \modulc{\En} dV } \leq 1.
\end{equation}
In the following, when using the DGF given by (\ref{Green}), one should keep in mind that $\vec{E}_n$ has to be replaced by $\En$.

\subsection{Universal statistics of modes in lossy chaotic RCs}
\label{lossy RC}

\begin{figure}
  \center
  \hspace*{-0.3 cm} \mbox{
    \raisebox{4.2cm}[0cm][0cm]{(a)}\hspace*{-1.2 em}
     {\resizebox{0.49\columnwidth}{!}{\includegraphics{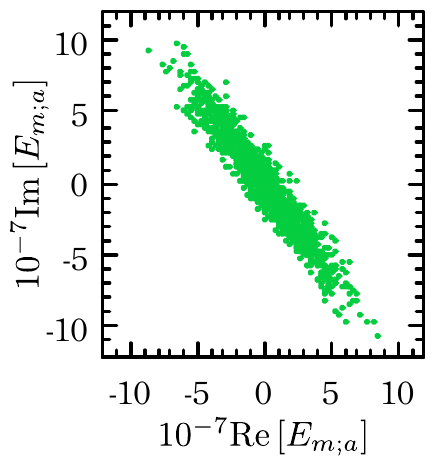}}}
    \raisebox{4.2cm}[0cm][0cm]{(b)}\hspace*{-1.2 em}
      {\resizebox{0.49\columnwidth}{!}{\includegraphics{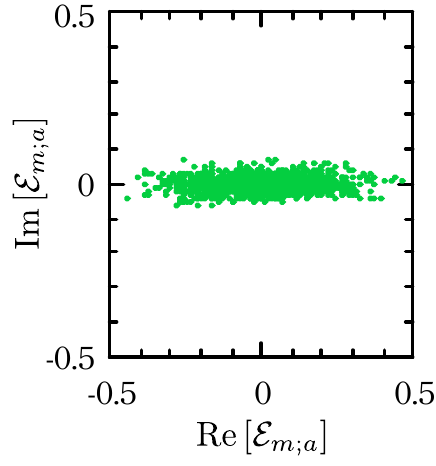}}}
  }
\caption{\label{fig_norm} (color online)
Imaginary vs.\ real part of all Cartesian components of the electric field before (a) and after (b) normalization by Eq.~\eqref{norma_vec}.}
\end{figure}

As emphasized in Ref.~\citep{GrosWamot2014}, the ideal 3D chaotic cavity is given by a fully asymmetric room without any parallel walls and with defocusing parts (focusing parts can also be used, with restrictions regarding the centers of curvature \cite{Bunimovich1998,Berry1981}).
Therefore, using a parallelepipedic room to ensure a homogeneous distribution of energy is not an optimal option as acknowledged by the community of room acoustics (see Ref.~\cite{Mortessagne1993} and references therein), and more recently in the EMC community \cite{coro}.
Since we wish to address the physical situations encountered in reverberation chambers, we start from a parallelepipedic cavity and introduce simple modifications of the boundary. The reverberation chambers are depicted in detail in Fig.~\ref{fig:lossy:cav}.
Moving the hemisphere along the ceiling will provide the necessary stirring in order to generate statistically independent configurations of the cavity (statistical ensemble).
None of the caps are centered and both penetrate inside the cavity to a maximum length of $15$\,cm.
The chaotic cavity thus obtained is a 3D realization of a dispersing billiard, the well-known Sinai billiard \cite{Alt1997,Dorr1998}.
By such modifications, the usable volume of the cavity is not significantly reduced, and simultaneously, as demonstrated in Ref.~\cite{GrosWamot2014}, the cavity thus obtained is an excellent realisation of a chaotic RC verifying all expected universal statistical properties of closed chaotic systems.

\begin{figure}
  \includegraphics{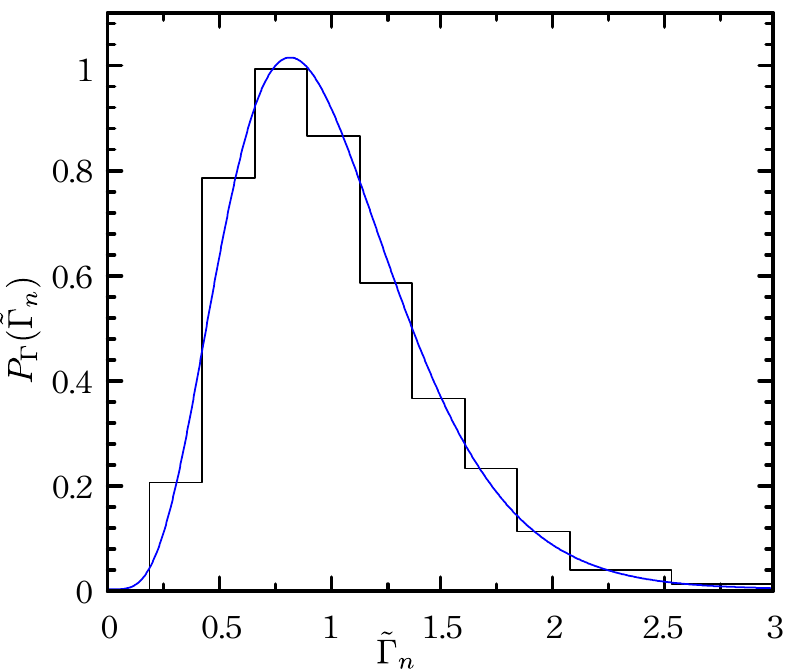}
\caption{\label{fig_gamma}(color online)
Comparison between empirical distributions of the normalized widths (black histogram) obtained in the \emph{cavity 1} shown in Fig.~\ref{fig:lossy:cav} where $d=0.43$ and formula (\ref{P_de_Gamma}) with $M=2\Gamma^2/\textrm{Var}(\Gamma_n)=11$ (continuous blue curve). The values of $\sqrt{\textrm{Var}(\Gamma_n)}/\Delta $ is 0.19.}
\end{figure}

In the present study, we concentrate on the range of frequencies around 1 GHz (close to the above defined LUF, near the 210th eigenstate).
For cavities with such dimensions, where losses are due only to the finite conductivity of the walls, the mean quality factor $Q$ would be of the order of $10^4-10^5$.
However, in practice, the latter is rather of the order of $10^3$ due to leakage through antennas or defects in the walls or sundry objects under test.
For the sake of simplicity, in our model we mimic the reduced quality factor by introducing sub-surfaces (colored patches in Fig.~\ref{fig:lossy:cav}) of reduced conductivity $\sigma_{\textrm{abs}}^c$ distributed on the walls. The total area of these sub-surfaces is $S_{\textrm{abs}}$.
The conductivity of copper ($\sigma^c_{Cu}=5.7 \times 10^7 $ S/m) is attributed to the remaining surface (in gray in Fig.~\ref{fig:lossy:cav}).
The values of $\sigma_{\textrm{abs}}^c$ are chosen so that $Q$ takes on realistic values comprising between 1500 and 2000.
While modifying $S_{\textrm{abs}}$ (passing from the configuration with the three red patches, now indicated as \emph{cavity 1}, to the configuration with six patches, indicated as \emph{cavity 2}), $Q$ is approximately kept constant by adjusting $\sigma_{\textrm{abs}}^c \propto Q^2 S^2_{\textrm{abs}}$ \cite{jackson_classical_1999,stoeckmann1999quantum,Hill_2}.
In our model, typical values of $\sigma_{\textrm{abs}}^c$ are $\sim 10^{-7}\sigma^c_{Cu} $. With such a huge ratio, only the non-copper parts of the walls contribute significantly to the losses.
The eigenvalues and eigenmodes were obtained by solving the vectorial Helmholtz equation with impedance boundary conditions with the help of the commercial software Comsol.
We noticed that the solver gives an arbitrary global phase to the vectorial electric field $\vec{E}_n$ associated with the $n$th resonance [Fig.~\ref{fig_norm}(a)].
The phase is identical for the three Cartesian components of the electric field but changes from mode to mode.
The effect of the normalization \eqref{norma_vec} is shown in Fig.~\ref{fig_norm}(b), where we can check that the real and imaginary parts of Cartesian components of $\En$, viewed as spatially distributed random variables, are thus made independent.
The lossy patches give rise to the $W$ term in the effective Hamiltonian as they are opening the system to the `environment' (for details see Refs.~\cite{bar05a,Barthelemy2005_2}).

For 30 different configurations obtained by moving the hemisphere along the ceiling, we computed 50 eigenmodes around the 200th resonance.
For a modal overlap $d\sim 0.45$, the values of $\sqrt{\textrm{Var}(\Gamma_n)}/\Delta $ are of the order of 0.15 for both cavities, so that a weak coupling assumption is reasonable. Indeed, a good agreement between the empirical distributions of the widths and formula (\ref{P_de_Gamma}) with $M=2\, {\Gamma}^2/\textrm{Var}(\Gamma_n)$ is obtained (see Fig.~\ref{fig_gamma}).
In Fig.~\ref{fig_gamma} only results concerning \emph{cavity 1} are shown, those concerning \emph{cavity 2} lead to the same conclusion. Indeed in this cavity, the parameter $\sqrt{\textrm{Var}(\Gamma_n)}/\Delta =0.13 $ and the distribution of the widths is in perfect agreement with the distribution (\ref{P_de_Gamma}) where $M$ is evaluated through $M=2\, {\Gamma}^2/\textrm{Var}(\Gamma_n)=23$.

\begin{figure}
  \raisebox{5.5cm}[0cm][0cm]{(a)}\hspace*{-1em}
    {\resizebox{\columnwidth}{!}{\includegraphics{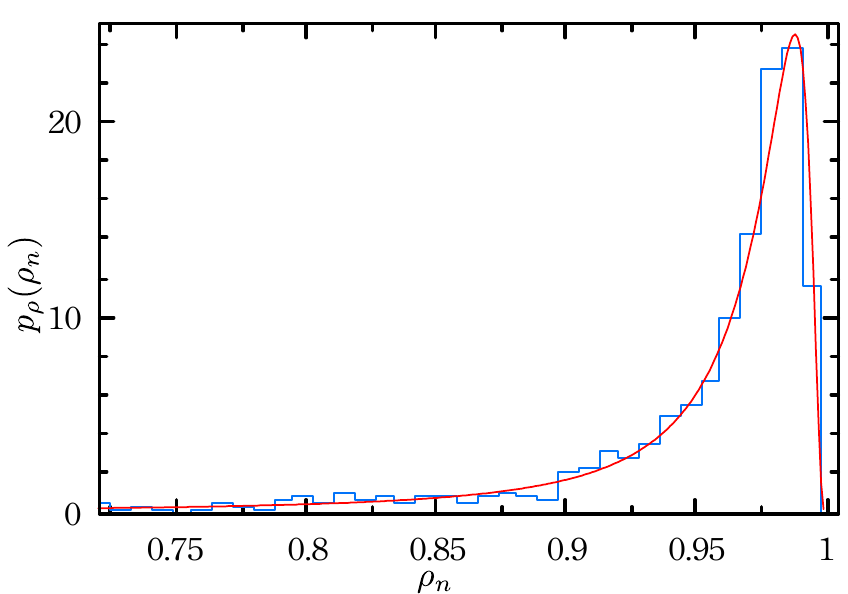}}}
  \raisebox{6.8cm}[0cm][0cm]{(b)}\hspace*{-1em}
    {\resizebox{\columnwidth}{!}{\includegraphics{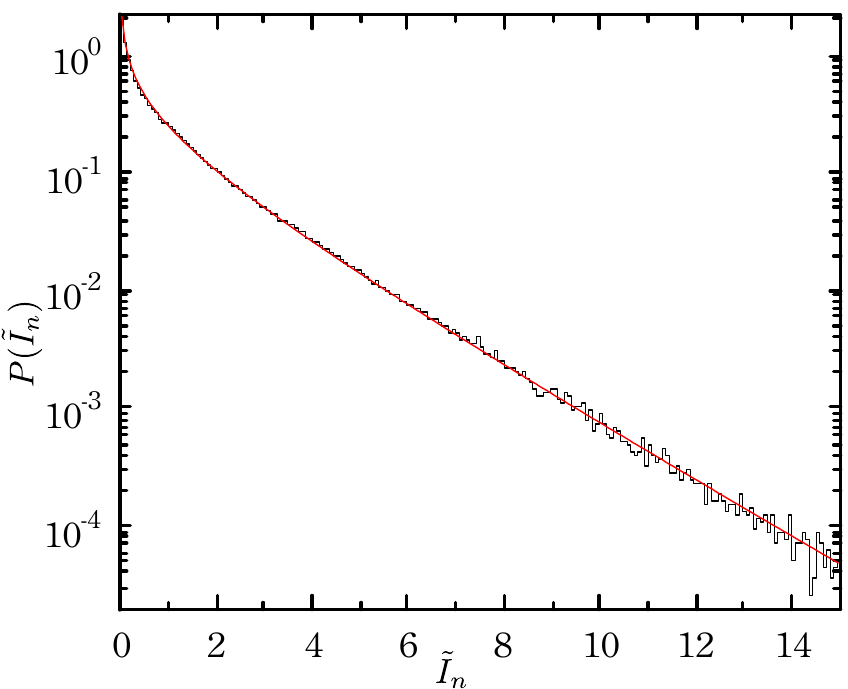}}}
\caption{\label{fig_rho_n_In_3patch}(color online)
In \emph{cavity 1}, (a) the distribution of the phase rigidity $\rho_n$ (blue histogram) of the $50 \times 30$ eigenfields is compared to the analytical perturbative $p_{\rho;\lbrace M,d \rbrace}$ (red curve) with $d$=0.43 and $M=11$. (b) The distribution of the normalized intensity (black histogram) of the three components of the same set of eigenfields is compared to expression (\ref{P_de_In_global}) (red curve) using the analytical $p_{\rho;\lbrace M,d \rbrace}$ shown in panel (a).}
\end{figure}

The next step is to recover the distribution of the phase rigidity.
It will now be shown that the relation (\ref{q2_et_rho}) between $\rho_n$ and $q^2_n$ permits to properly define an extended complexness parameter which, for the case of vectorial fields in chaotic cavities, where the modes are expected to be homogenously distributed and isotropic, must follow the same distribution as in the scalar case. Indeed, by combining Eqs.~\eqref{q2_et_rho} and \eqref{modul_rho_n_vec}, one has
\begin{equation}
q^2_n =\frac{ \av{\modulc{\En}}-1 }{ \av{\modulc{\En}}+1} \label{defq2_rho}
\end{equation}
The bi-orthogonality conditions (\ref{norm_int}) yield
\begin{displaymath}
\left\{
\begin{array}{l}
\av{\Ree{\En}\ccdot\Ree{\En}}+\av{\Imm{\En}\ccdot\Imm{\En}}= \av{\modulc{\En}}\\
\av{\Ree{\En}\ccdot\Ree{\En}}-\av{\Imm{\En}\ccdot\Imm{\En}}=1\,,
\end{array}
\right.
\end{displaymath}
and one can deduce from Eq.~\eqref{defq2_rho} the following extended definition for the complexness parameter :
\begin{equation}\label{defq2}
  q^2_n=\frac{\av{\Imm{\En}\ccdot\Imm{\En}}}{\av{\Ree{\En}\ccdot\Ree{\En}}} \, .
\end{equation}
Since we are interested in chaotic cavities, we can assume that the eigenfields $\En$ are statistically isotropic so that the Cartesian components $\Enc{x}$,$\Enc{y}$,$\Enc{z}$ are independent and identically distributed random variables yielding
\begin{eqnarray}
 q^2_n&=\frac{\av{\Imm{\Enc{x}}^2+\Imm{\Enc{y}}^2+\Imm{\Enc{z}}^2}}{\av{\Ree{\Enc{x}}^2+\Ree{\Enc{y}}^2+\Ree{\Enc{z}}^2}} \nonumber \\
&= \frac{\av{\Imm{U}^2}}{\av{\Ree{U}^2}}\,, \label{defq2_b}
\end{eqnarray}
where $U$ follows the same distribution as the scalar eigenfields $\Psi_n$. Hence the extended complexness parameter (and the corresponding phase rigidity), as well as the normalized intensities of the Cartesian components of the eigenfields, must obey the same distribution laws as for the scalar case [Eqs.~\eqref{p_de_x} and \eqref{P_de_In_global}].
This is illustrated in Fig.~\ref{fig_rho_n_In_3patch} for \emph{cavity 1}, where the distribution of $\rho_n$ (blue histogram) is computed for all the eigenfields mentioned above and compared to the perturbative $p_{\rho;\lbrace M,d \rbrace}$ (red curve) with $d=0.43$ and $M=11$ as deduced from the nearest integer of formula \eqref{eq:M:gamma}.
In the bottom part of the figure, the distribution of the normalized intensity (histogram) of the three components of all eigenfields is compared to expression (\ref{P_de_In_global}) (red curve) with the above mentioned analytical $p_{\rho;\lbrace M,d \rbrace}$.
In this case where the weak coupling limit is ensured, the agreement between the empirical distributions and the RMT predictions is remarkable.

\begin{figure}
\resizebox{\columnwidth}{!}{\includegraphics{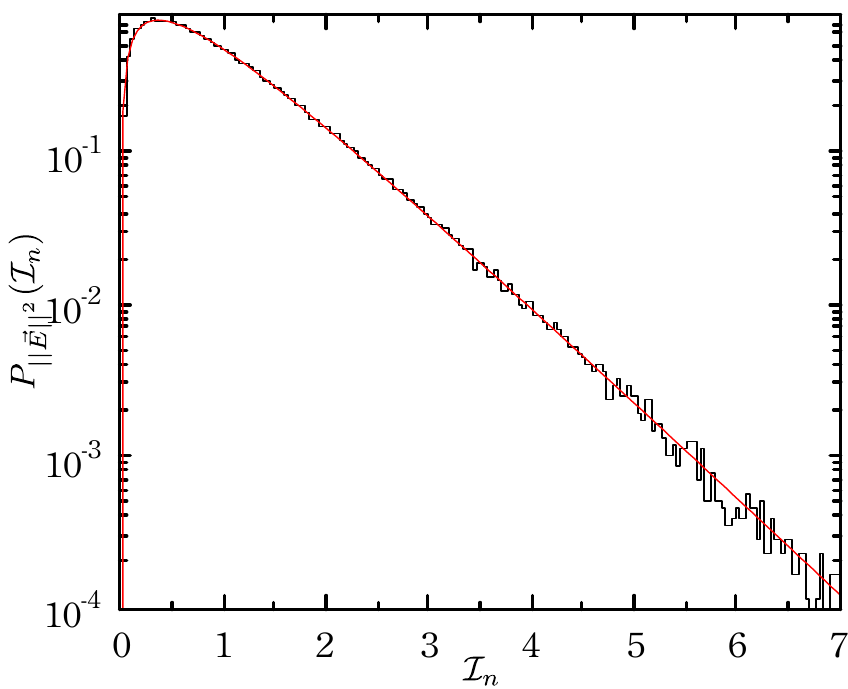}}
\caption{\label{fig:In:vec:cav}(color online)
The distribution of the normalized modulus squared of the 1500 computed eigenfields of \emph{cavity 1} (black histogram) is compared to expression \eqref{eq:P_It:vec:int} (red curve) with the analytical $p_{\rho;\lbrace M;d\rbrace}$ shown in Fig.~\ref{fig_rho_n_In_3patch}(a).}
\end{figure}

We have thus demonstrated that, thanks to an appropriate definition of the phase rigidity, the Cartesian components of eigenfields of vectorial chaotic systems and eigenvectors of random $\Heff$ are statistically equivalent. However, some typical quantities related with a vectorial field problem such as the distribution of the normalized modulus squared of the vector field, defined for the eigenfield $\En$ by
\begin{equation}\label{eq:modulus:squarre}
  \Icn=\frac{\Enc{x}\Enc{x}^*+\Enc{y}\Enc{y}^*+\Enc{z}\Enc{z}^*}{\av{\Enc{x}\Enc{x}^*+\Enc{y}\Enc{y}^*+\Enc{z}\Enc{z}^*}}=\frac{\En\ccdot\En^*}{\av{\En\En^*}}\,,
\end{equation}
have no direct equivalent in RMT formalism.
Nevertheless, the knowledge of statistical properties of Cartesian components brings the key elements to derive the distribution of $\Icn$.
Indeed, the numerator of \eqref{eq:modulus:squarre} can be written as
\begin{equation}
  \En\ccdot\En^*=\Ree{\En}\cdot \Ree{\En}+\Imm{\En}\cdot\Imm{\En},
\end{equation}
and, as shown above, in chaotic RCs due to the statistical uniformity and isotropy of the EM field, $\Ree{\En}\cdot \Ree{\En}$ and $\Imm{\En}\cdot\Imm{\En}$ are two independent random variables, both are equivalent to the sum of the square of three identically distributed Gaussian random variables.
But the Gaussian random variables corresponding to $\Ree{\En}\cdot \Ree{\En}$ and those corresponding to $\Imm{\En}\cdot\Imm{\En}$ are not necessarily the same. Hence $\Ree{\En}\cdot \Ree{\En}$ and $\Imm{\En}\cdot\Imm{\En}$ are two $\chi^2_3$ random variables whose ratio of their mean values is fixed by the value of the complexness parameter $q_n^2$ [see Eq.~\eqref{defq2}].
 Consequently, for a given eigenfield $\En$, the distribution of its normalized intensities $\Icn$ is given by \cite{Bausch2013}
\begin{equation}\label{eq:P_It:vec}
P_\Ic(\Icn;\rho)=
\frac{9\,\Icn}{\sqrt{\abs{\rho}^2-\abs{\rho}^4}}\exp\left(-\frac{3 \Icn}{1-\abs{\rho} ^2}\right)
\textrm{I}_1\left(\frac{3 \Icn\abs{\rho} }{1-\abs{\rho} ^2}\right)
\end{equation}
with $\textrm{I}_1$ the modified Bessel function of the first kind and $\rho=\rho_n$ the phase rigidity of $\En$.
Note that the difference between this equation and Eq.~\eqref{P_de_In} stems from the different degrees of freedom in the statistical components.
Therefore, the distribution of the normalized modulus squared of an ensemble of eigenfields follows
\begin{equation}\label{eq:P_It:vec:int}
  \PIvec(x)=\int p_\rho(\rho) P_\Ic(\Ic(x); \rho) d\rho,
\end{equation}
where $p_\rho=p_{\rho;\lbrace M;d\rbrace}$ in the case of the weak coupling regime.
The reader's attention should be drawn to the fact when $\abs{\rho}$ tends to 1, $P_\Ic\to \chi^2_3$ (which is expected for lossless chaotic RCs) and when $\rho$ tend to 0, $P_\Ic\to \chi^2_6$ (which is expected for fully open chaotic RCs).
The validity of the above derivation is verified in Fig.~\ref{fig:In:vec:cav} where the distribution of $\Icn$ of the previously mentioned set of eigenfields of \emph{cavity 1} (black histogram) is compared to distribution \eqref{eq:P_It:vec:int} (red curve), where $p_\rho$ is replaced by the perturbative $p_{\rho;\lbrace M;d\rbrace}$ shown in Fig.~\ref{fig_rho_n_In_3patch}(a).

\subsection{Effectiveness of random matrix to predict statistics of the response in chaotic RCs}

\begin{figure}
\center
\resizebox{\columnwidth}{!}{\includegraphics {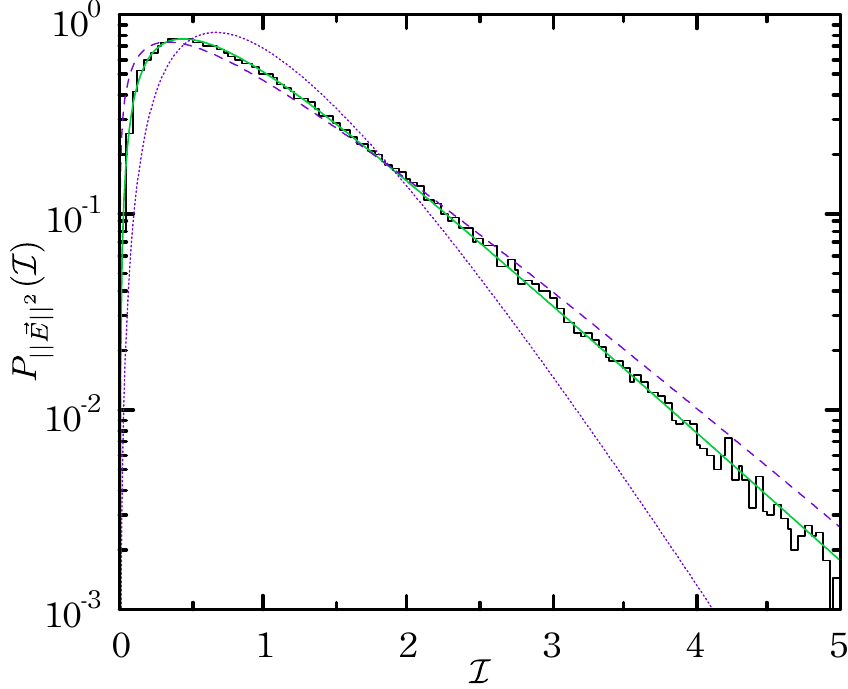}}
\caption{\label{fig:P:de:I:vec}(color online)
Comparison of the distribution of the normalized modulus squared of an ensemble of EM responses of \emph{cavity 2} (black histogram) and the distribution \eqref{eq:P:I:vec:rep} where $P_\rho$ is the synthetic distribution obtained via RMT response simulation with $M=23$ and $d=0.43$ (green continuous curve).
For the sake of comparison, the $\chi^2_3$ distribution, expected for closed systems, and $\chi^2_6$, expected for fully open systems, are represented by the dashed and dotted violet lines, respectively.}
\end{figure}

In this last section, we are interested in the distribution of the normalized intensity of Cartesian components or equivalently in the distribution of the normalized modulus squared of an ensemble of EM responses in chaotic RCs.
Originally, the distribution (\ref{P_de_In}) was proposed by Pnini and Shapiro to describe the statistics of the wave function of open disordered or chaotic systems \cite{Pnini}.
For a given configuration and at a given excitation frequency, the phase rigidity of the wave function is thus required.
In the case of well isolated resonances, the phase rigidity of the wave function given by the response (\ref{Green_mq_exp}) is determined by the values of the phase rigidity of the individual resonance state and is really meaningful at the central frequencies of the resonances.
Obviously, when the modal overlap comes into play, the phase rigidity of the response is built upon several different states which contribute according to their spatial overlaps with the source and to how far their central frequencies are from the excitation frequency.
Therefore the distribution of the phase rigidity of the response is generally different from the corresponding quantity for the individual eigenfunctions, as already emphasized in Ref.~\citep{bulg06b}.
Consequently, for an ensemble of EM responses $\lbrace \E \rbrace$, the distribution of the normalized intensity of each Cartesian component and the distribution of the normalized intensity of the response are expected to be given by
\begin{equation}\label{eq:P:I:rep}
  P(\tilde{I})=\int P_\rho(\rho) P(\tilde{I}; \rho) d\rho\,,
\end{equation}
\begin{equation}\label{eq:P:I:vec:rep}
  \PIvec(x)=\int P_\rho(\rho) P_\Ic(\Ic(x); \rho) d\rho\,,
\end{equation}
respectively, where $\rho=\av{\E\cdot\E} / \av{\,\normec{\E}}$ and $P_\rho$ is, to our knowledge, not known analytically and only empirically accessible through a statistical sample of values of $\rho$.
However, if the number of effective coupling channels $M$ and their coupling strength $\kappa$ are known, the statistical behavior of any open chaotic system can be simulated through random $\Heff$ simulations.
Especially, the distribution of the phase rigidity of an ensemble of responses of a realistic chaotic system should correspond with the phase rigidity
distribution of an ensemble of random matrix responses \eqref{Green_mq_exp} built upon the eigenvectors and eigenvalues of random $\Heff$, where $M$ and $\kappa$ are those of the corresponding realistic system.
This synthetic distribution of $\rho$ produced by RMT simulations can then replace $P_\rho$ in Eqs.~\eqref{eq:P:I:rep} and \eqref{eq:P:I:vec:rep} and the resulting distributions should predict respectively the intensity distribution of Cartesian components and the
distribution of normalized modulus square of an ensemble of EM responses of chaotic RCs.
We now demonstrate the relevance of the above surmise in the case of an ensemble of response computed in the chaotic
electromagnetic reverberation chamber \emph{cavity 2}. This ensemble is obtained by
varying the configuration given here by the threesome: excitation frequency, polarization of the source and position of the
hemisphere. More precisely, we computed 50 eigenmodes of \emph{cavity 2} around the 200th
resonance for 30 positions of the hemisphere, and built the response using Eq.~\eqref{eq:E:avec:tgd} (evaluated at 16 different points inside the cavity) by means of the DGF \eqref{Green}, for 300 equidistant excitation frequencies comprised in the interval $\left[ 969.4 \,\textrm{MHz},1024\,\textrm{MHz}\right]$.
For the results presented below, the average modal overlap is $d=0.43$ corresponding to a mean quality factor $Q=1500$.
As mentioned above, in this frequency range, a weak coupling regime can be assumed for \emph{cavity 2}.
Thus the number of coupling channels as well as the coupling strength can be estimated thanks to Eq.~\eqref{eq:M:gamma} $M=2\Gamma^2/\textrm{Var}(\Gamma_n)=23$ and Eq.~\eqref{kappa_d} $\kappa=\pi d/(2M)=0.03$.
In Fig.~\ref{fig:P:de:I:vec}, the distribution of the normalized modulus square of the above mentioned ensemble of EM responses is compared with the distribution \eqref{eq:P:I:vec:rep} where $P_\rho$ is the synthetic distribution obtained via RMT response simulation with $M=23$ and $d=0.43$.
A good agreement between histogram and RMT prediction is observed. In contrast, deviations from the predictions for the closed and fully open systems are found.
Thus we find universal statistics for all fully-chaotic partly-open reverberation chambers with the same number of effective channels and modal overlap.

\section{Conclusion}
Starting from the effective Hamiltonian $\Heff$ formalism including random matrix theory, we can derive the distribution of intensities for scalar waves in the weak coupling limit.
The only parameters entering our model are the modal overlap and the number of open channels.
For weak coupling (corresponding to small or moderate modal overlap) we find good agreement between random matrix eigenvectors and our theoretical predictions, whereas, in the range of strong coupling, deviations are found.
The discrepancies are essentially due to the deviations in the distribution of the phase rigidity.
For further comparison, we calculated the electric field distribution inside a lossy chaotic reverberation chamber.
By comparing the thus obtained width distributions with the $\chi^2$ distribution expected in the weak coupling regime, we extract the only two parameters, namely the modal overlap and the number of open channels.
Concerning the vectorial electric eigenfield, a good agreement is seen for the distributions of (i) the phase rigidity $\abs{\rho_n}$, (ii) the individual Cartesian component intensities $\tilde{I}_n$ as well as (iii) the field intensity $\Ic$.
Finally, we derive the distribution for the response, typically the quantity of interest in reverberation chambers used for electromagnetic compatibility. The obtained distribution corresponds well with the theoretical one, where the synthetic distribution of the phase rigidity obtained from random matrix numerics is used. We would like to emphasize that, in all cases, the only parameters are the modal overlap and the number of channels, which are known a priori.

\begin{acknowledgements}
We acknowledge financial support by the ANR CAOREV and are thankful to E.~Richalot, P.~Besnier and D.~Savin for fruitful discussions.
\end{acknowledgements}

\bibliographystyle{apsrev4-1}
\bibliography{LCEMRC}

\end{document}